\documentstyle[amssymb]{article}

\begin{document}

\title{An equivalence of two mass generation mechanisms for gauge fields}

\author{A. Sevostyanov \\
Department of Mathematical Sciences \\
University of Aberdeen  \\ Aberdeen AB24 3UE, United Kingdom \\
e-mail: seva@maths.abdn.ac.uk}

\maketitle

\begin{abstract}
Two mass generation mechanisms for gauge theories are studied. It
is proved that in the abelian case the topological mass generation
mechanism introduced in Refs. \cite{BCS,HL,L}  is equivalent to
the mass generation mechanism defined in Refs. \cite{C,S} with the
help of ``localization'' of a nonlocal gauge invariant action. In
the nonabelian case the former mechanism is known to generate a
unitary renormalizable quantum field theory describing a massive
vector field.
\end{abstract}

\vskip 1cm \noindent {\em PACS numbers:} 11.15.–q.

\noindent {\em Keywords and phrases:} Yang-Mills field, mass
generation.

\section*{Introduction}

In the last two decades several mass generation mechanisms for
nonabelian gauge fields were suggested (see Ref. \cite{Jw} for
discussion of these mechanisms in three--dimensional case). In the
framework of perturbation theory these mechanisms are expected to
provide a new theoretical background for describing the
electroweak sector of the Standard Model in such a way that the
unobserved Higgs boson does not appear in the physical spectrum.
On the other hand the problem of finding mass terms in gauge
theories is motivated by nonperturbative Quantum Chromodynamics.

In this paper we study two mass generation mechanisms for
nonabelian gauge theories introduced in Refs. \cite{BCS,HL,L} and
Refs. \cite{C,S} in case of four--dimensional space--time. In
Refs. \cite{C,S} a new classical gauge invariant nonlocal
Lagrangian generating a local quantum field theory was
constructed. This phenomenon is similar to that for the
Faddeev-Popov determinant in case of the quantized Yang-Mills
field. Recall that being a priori nonlocal quantity the
Faddeev-Popov determinant can be made local by introducing
additional anticommuting ghost fields and applying a formula for
Gaussian integrals over Grassmann variables. Similarly, in case of
the nonlocal Lagrangian suggested in Refs. \cite{C,S} one can
introduce extra ghost fields, both bosonic and fermionic, and make
the expression for the generating function of the Green functions
local using tricks with Gaussian integrals. In the local
expression for the Green functions the ``localized'' Lagrangian
containing extra ghost fields should be used instead of the
original one. It was shown in Refs. \cite{C,C1,S1} that the
corresponding ``localized'' Lagrangian containing extra ghost
fields is renormalizable. When the coupling constant vanishes the
nonlocal Lagrangian, in a certain gauge, is reduced to that of
several copies of the massive vector field.

In fact the ``localized'' Lagrangian describes the gauge field $A$
coupled to an antisymmetric $(2,0)$-type tensor potential $\Phi$
via the topological term ${\mathrm tr}~(*\Phi\wedge F)$ with a
coupling constant $m$ of mass dimension one, $F$ being the
curvature of $A$, and $*$ is the Hodge star operator (here and
below we assume that the tensor fields take values in a compact
Lie algebra $\mathfrak g$, and ${\mathrm tr}$ is an invariant
scalar product in that Lie algebra). The ``localized'' Lagrangian
also contains a gauge invariant kinetic term for $\Phi$ and gauge
invariant kinetic terms for the fermionic ghost fields.

In this paper we show that there are some hidden symmetries for
the corresponding abelian ``localized'' Lagrangian, and using
these symmetries one can define a physical sector of the theory in
a consistent way. In the physical sector the abelian ``localized''
Lagrangian describes $\mathfrak g$--valued massive vector field.

There is another mass generation mechanism for nonabelian gauge
fields for which the corresponding Lagrangian is constructed with
the help of an antisymmetric $\mathfrak g$--valued $(2,0)$-type
tensor potential $B$ coupled to the gauge field $A$ via the
topological term ${\mathrm tr}~(B\wedge F)$. This mechanism was
suggested in Refs. \cite{BCS,HL,L}. Using BRST cohomology
technique one can prove that the corresponding nonabelian gauge
field theory is unitary and renormalizable (see Refs. \cite{HL,L1}
and Ref. \cite{L2}). In the physical sector the theory describes
the massive $\mathfrak g$--valued vector field. So that there are
some similarities between constructions suggested in Refs.
\cite{BCS,HL,L} and Refs. \cite{C,S}.

In this paper we prove that in the abelian case the massive gauge
theories constructed in Refs. \cite{BCS,HL,L} and Refs. \cite{C,S}
are equivalent. Beside of the gauge symmetry the action for the
abelian theory defined in Refs. \cite{BCS,HL,L} has also a vector
symmetry, and the abelian version of the action introduced in
Refs. \cite{C,S} is a gauge fixed version of the former one, with
respect to the vector symmetry. So that in both cases the physical
sector can be described with the help of the BRST cohomology
corresponding to the gauge and the vector symmetries.

This paper is organized as follows. In Section \ref{sect1} we
recall the main construction of Refs. \cite{C,S} and fix the
notation used throughout of the paper. In Section \ref{hamform}
the Hamiltonian formulation for the nonabelian theory suggested in
Refs. \cite{C,S} is introduced. Then we study the corresponding
unperturbed abelian theory in Section \ref{unp}. In particular, we
find a canonical form for the corresponding unperturbed quadratic
Hamiltonian and study the symmetries of this Hamiltonian. It turns
out that there are some hidden first class constraints for the
unperturbed Hamiltonian. These constraints allow to reduce the
number of physical degrees of freedom. In Section \ref{SQ} we
quantize the unperturbed system and show that one can define a
physical sector for the quantized theory. The physical sector
describes the quantized $\mathfrak g$--valued massive vector
field. In Section \ref{unit} we compare the actions defined in
Refs. \cite{BCS,HL,L} and Refs. \cite{C,S} in the abelian case. We
prove that the abelian version of the action introduced in Refs.
\cite{C,S} is a gauge fixed version of the abelian action defined
in Refs. \cite{BCS,HL,L}. Following Refs. \cite{HL,L1} we also
define the BRST cohomology which can be used to describe the
physical sector for both theories.

\section{Recollection}\label{sect1}

In this section we recall the definition of the action introduced
in Refs. \cite{C,S} for describing nonabelian massive gauge
fields. First we fix the notation as in Ref. \cite{IZ}. Let $G$ be
a compact simple Lie group, $\mathfrak g$ its Lie algebra with the
commutator denoted by $[\cdot, \cdot ]$. We fix a nondegenerate
invariant under the adjoint action scalar product on $\mathfrak g$
denoted by $\textrm{tr}$ (for instance, one can take the trace of
the composition of the elements of $\mathfrak g$ acting in the
adjoint representation). Let $t^a$, $a=1,\ldots
,\textrm{dim}{\mathfrak g}$ be a linear basis of $\mathfrak g$
normalized in such a way that
$\textrm{tr}(t^at^b)=-\frac{1}{2}\delta^{ab}$.

We denote by $A_\mu$ the $\mathfrak g$-valued gauge field
(connection on the Minkowski space),
$$
A_\mu=A_\mu^at^a.
$$
Let $D_\mu$ be the associated covariant derivative,
$$
D_\mu=\partial_\mu-\textrm{g}A_\mu,
$$
where $\textrm{g}$ is a coupling constant, and $F_{\mu\nu}$ the
strength tensor (curvature) of $A_\mu$,
$$
 F_{\mu\nu}=\partial_\mu
A_\nu - \partial_\nu A_\mu -\textrm{g}[A_\mu,A_\nu].
$$

We shall also need a covariant d'Alambert operator $\square_A$
associated to the gauge field $A_\mu$,
$$
\square_A=D_\mu D^\mu.
$$
The covariant d'Alambert operator can be applied to any tensor
field defined on the Minkowski space and taking values in a
representation space of the Lie algebra $\mathfrak g$, the
$\mathfrak g$-valued gauge field $A_\mu$ acts on the tensor field
according to that representation.

Finally recall that the gauge group of $G$-valued functions $g(x)$
defined on the Minkowski space acts on the gauge field $A_\mu$ by
\begin{equation}\label{gaugen}
A_\mu \mapsto \frac{1}{\textrm{g}}(\partial_\mu g)g^{-1}+gA_\mu
g^{-1}.
\end{equation}
The corresponding transformation laws for the covariant derivative
and the strength tensor are
\begin{eqnarray}
D_\mu \mapsto gD_\mu g^{-1}, \label{trcov}\\
F_{\mu\nu} \mapsto gF_{\mu\nu}g^{-1}. \label{trf}
\end{eqnarray}
Formula (\ref{trcov}) implies that the covariant d'Alambert
operator is transformed under gauge action (\ref{gaugen}) as
follows
\begin{equation}\label{trdal}
\square_A \mapsto g\square_A g^{-1}.
\end{equation}
In the last formula it is assumed that the gauge group acts on
tensor fields according to the representation of the group $G$
induced by that of the Lie algebra $\mathfrak g$.

The ``localized'' action for the massive gauge field introduced in
Refs. \cite{C,S} can defined by the formula
\begin{eqnarray}\label{act1}
S=\int  \textrm{tr} \left(
\frac{1}{2}F_{\mu\nu}F^{\mu\nu}-\frac{1}{8}(\square_A\Phi_{\mu\nu})\Phi^{\mu\nu}
+\frac{m}{2}\Phi_{\mu\nu}F^{\mu\nu} -2i\sum_{i=1}^3
\overline{\eta}_i (\square_A \eta_i)\right)  d^4x.
\end{eqnarray}
Here $\Phi_{\mu\nu}$ is a skew--symmetric (2,0)-type tensor field
in the adjoint representation of $\mathfrak g$; $\eta_i,
\overline{\eta}_i$, $i=1,2,3$ are pairs of anticommuting scalar
ghost fields in the adjoint representation of $\mathfrak g$; they
satisfy the following reality conditions: $\eta_i^*= {\eta}_i$,
$\overline{\eta}_i^*= \overline{\eta}_i$. In formula (\ref{act1})
$\textrm{g}$ should be regarded as a coupling constant and $m$ is
a mass parameter. From (\ref{trcov}), (\ref{trf}) and
(\ref{trdal}) it follows that action (\ref{act1}) is invariant
under gauge transformations (\ref{gaugen}).

To define the Green functions corresponding to the gauge invariant
action $S$ we have to add to action (\ref{act1}) another term
$S^{gf}$ containing a gauge fixing condition and the corresponding
Faddeev-Popov operator. As it was observed in Ref. \cite{S} the
most convenient choice of $S^{gf}$ is
$$
S^{gf}=\int \textrm{tr} \left(\partial^\mu A_\mu \partial^\nu
A_\nu + m^2\partial^\mu A_\mu (\square^{-1}_F
\partial^\nu A_\nu)-2i\overline{\eta} (
\partial^\mu D_\mu
\eta) \right)d^4x,
$$
where $\square^{-1}_F$ is the operator inverse to d'Alambert
operator with radiation boundary conditions, $\eta,
\overline{\eta}$ is a pair of anticommuting scalar ghost fields in
the adjoint representation of $\mathfrak g$; they satisfy the
following reality conditions: $\eta^*= {\eta}$,
$\overline{\eta}^*= \overline{\eta}$.

The total action $S'=S+S^{gf}$, that should be used in the
definition of the Green functions, takes the form
\begin{eqnarray}\label{actloc}
S'=\int  \textrm{tr} \Big(
\frac{1}{2}F_{\mu\nu}F^{\mu\nu}-\frac{1}{8}(\square_A\Phi_{\mu\nu})\Phi^{\mu\nu}
+\frac{m}{2}\Phi_{\mu\nu}F^{\mu\nu}-2i\sum_{i=1}^3
\overline{\eta}_i (\square_A \eta_i)+ \\
 +\partial^\mu A_\mu
\partial^\nu A_\nu + m^2\partial^\mu A_\mu (\square^{-1}_F
\partial^\nu A_\nu)-2i\overline{\eta} (
\partial^\mu D_\mu
\eta)\Big) d^4x.\nonumber
\end{eqnarray}

The generating function
$Z(J,I,\xi,\overline\xi,\xi^i,\overline\xi^i)$ of the Green
functions corresponding to (\ref{actloc}) is
\begin{eqnarray}
Z(J,I,\xi,\overline\xi,\xi^i,\overline\xi^i)=\int{\mathcal
D}(A_\mu){\mathcal D}(\Phi_{\mu \nu}){\mathcal D}(\eta){\mathcal
D}(\overline{\eta})\prod_{i=1}^3{\mathcal D}(\eta_i){\mathcal
D}(\overline{\eta}_i) \times \qquad \qquad \qquad \nonumber
\\
\times \exp \Big\{ i \int \textrm{tr} \Big(
\frac{1}{2}F_{\mu\nu}F^{\mu\nu}-\frac{1}{8}(\square_A\Phi_{\mu\nu})\Phi^{\mu\nu}
+\frac{m}{2}\Phi_{\mu\nu}F^{\mu\nu}-2J^\mu A_\mu -I^{\mu\nu}\Phi_{\mu\nu}-  \label{gener} \\
-2\overline\xi\eta-2\xi\overline\eta
-2\overline\xi^i\eta_i-2\xi^i\overline\eta_i -2i\overline{\eta} (
\partial^\mu D_\mu
\eta)-2i\sum_{i=1}^3 \overline{\eta}_i (\square_A
\eta_i)+ \nonumber \\
 +\partial^\mu A_\mu \partial^\nu A_\nu +
m^2\partial^\mu A_\mu (\square^{-1}_F
\partial^\nu A_\nu) \Big) d^4x \Big\}, \nonumber
\end{eqnarray}
where $J^\mu$, $I^{\mu \nu}$, $\xi$, $\overline\xi$, $\xi^i$,
$\overline{\xi}^i$, $i=1,2,3$ are the sources for the fields
$A_\mu$, $\Phi_{\mu \nu}$, $\eta$, $\overline\eta$, $\eta_i$,
$\overline{\eta}_i$, $i=1,2,3$, respectively.

Now consider the expression for the generating function
$Z(J)=Z(J,0,0,0,0,0)$ via a Feynman path integral,
\begin{eqnarray}\label{generJ}
Z(J)=\int{\mathcal D}(A_\mu){\mathcal D}(\Phi_{\mu \nu}){\mathcal
D}(\eta){\mathcal D}(\overline{\eta})\prod_{i=1}^3{\mathcal
D}(\eta_i){\mathcal D}(\overline{\eta}_i)\times \quad \qquad\qquad \qquad \qquad\qquad\qquad \\
\times\exp \left\{ i \int \textrm{tr} \left(
\frac{1}{2}F_{\mu\nu}F^{\mu\nu}-\frac{1}{8}(\square_A\Phi_{\mu\nu})\Phi^{\mu\nu}
+\frac{m}{2}\Phi_{\mu\nu}F^{\mu\nu}-2J^\mu A_\mu
-2i\overline{\eta} (
\partial^\mu D_\mu
\eta)- \right. \right. \nonumber \\
\left. \left. -2i\sum_{i=1}^3 \overline{\eta}_i (\square_A
\eta_i)+\partial^\mu A_\mu \partial^\nu A_\nu + m^2\partial^\mu
A_\mu (\square^{-1}_F
\partial^\nu A_\nu) \right) d^4x \right\} \nonumber.
\end{eqnarray}

Observe that in the r.h.s. of formula (\ref{generJ}) all the
integrals over the ghost fields are Gaussian. The Gaussian
integrals can be explicitly evaluated (see Refs. \cite{C,S} for
details). This yields
\begin{eqnarray}\label{act}
Z(J)=\int{\mathcal D}(A_\mu){\mathcal D}(\eta){\mathcal
D}(\overline{\eta})\exp \{ i \int [  \textrm{tr} \big(
\frac{1}{2}F_{\mu\nu}F^{\mu\nu}+\frac{m^2}{2}(\square^{-1}_AF_{\mu\nu})F^{\mu\nu}
\\
-2J^\mu A_\mu +\partial^\mu A_\mu \partial^\nu A_\nu +
m^2\partial^\mu A_\mu (\square^{-1}_F
\partial^\nu A_\nu)\big) ] d^4x \}{\rm det}(\partial^\mu D_\mu). \nonumber
\end{eqnarray}

The r.h.s. of (\ref{act}) looks like the generating function of
the Green functions for the Yang-Mills theory with an extra
nonlocal term, $\frac{m^2}{2}\textrm{tr}
\big((\square^{-1}_AF_{\mu\nu})F^{\mu\nu}\big)$, and in a
generalized gauge (see Ref. \cite{FS}, Ch 3, \S 3). The gauge
invariant action $S_m$ that appears in formula (\ref{act}),
\begin{eqnarray}\label{anonloc}
S_m=\int d^4x~\textrm{tr} \Big(
\frac{1}{2}F_{\mu\nu}F^{\mu\nu}+\frac{m^2}{2}(\square^{-1}_AF_{\mu\nu})F^{\mu\nu}
\Big),
\end{eqnarray}
is not local. But the generating function $Z(J)$ for this action
is equal to that for local action (\ref{act1}). This phenomenon
was observed in Refs. \cite{C,S}.

In case when the coupling constant g vanishes action
(\ref{anonloc}) takes the form
\begin{equation}\label{anonloc1}
S_m^0=\int d^4x~\textrm{tr} \Big(
\frac{1}{2}F_{\mu\nu}F^{\mu\nu}+\frac{m^2}{2}(\square^{-1}F_{\mu\nu})F^{\mu\nu}
\Big),~F_{\mu\nu}=\partial_\mu A_\nu -
\partial_\nu A_\mu.
\end{equation}
After imposing the Lorentz gauge fixing condition $\partial^\mu
A_\mu=0$ action (\ref{anonloc1}) coincides with the action of the
$\mathfrak g$-valued massive vector field (see Ref. \cite{S} for
details),
$$
S_m^0|_{\partial^\mu A_\mu=0}=\int d^4x~\textrm{tr} \Big(
\frac{1}{2}F_{\mu\nu}F^{\mu\nu}-m^2A_{\mu}A^{\mu}\Big),~F_{\mu\nu}=\partial_\mu
A_\nu -
\partial_\nu A_\mu.
$$

\section{Hamiltonian formulation}\label{hamform}

In this section we study the dynamical properties of the system
described by the Lagrangian $L$ that appears in formula
(\ref{actloc}),
\begin{equation}
L=\int  \textrm{tr} \left(
\frac{1}{2}F_{\mu\nu}F^{\mu\nu}-\frac{1}{8}(\square_A\Phi_{\mu\nu})\Phi^{\mu\nu}
+\frac{m}{2}\Phi_{\mu\nu}F^{\mu\nu} \label{actloc2}
-2i\sum_{i=1}^3 \overline{\eta}_i (\square_A \eta_i)\right)  d^3x.
\end{equation}

We start with the Hamiltonian formulation for the dynamical system
generated by Lagrangian (\ref{actloc2}). For the needs of the
Hamiltonian formulation we split the coordinates on the Minkowski
space into spatial and time components,
$x=(x^0,x^1,x^2,x^3)=(t,{\bf x})$, ${\bf x}=(x^1,x^2,x^3)$. We
shall also write $d^3x=dx^1dx^2dx^3$, and $\cdot$ will stand for
the scalar product in three-dimensional Euclidean space or
Minkowski space. For any $\mathfrak g$--valued quantity $X$ the
superscript $a$ will indicate the $a$-th component of $X$,
$X=X^at^a$; the Latin indexes will always take values $1,2,3$,
i.e. $i,j,k=1,2,3$; and summations over all repeated indexes will
be assumed.

In order to find the Hamiltonian formulation for the system
associated to the Lagrangian $L$ we rewrite $L$ in the following
form
\begin{eqnarray}\label{leg}
L=\int\left( p_i^a\partial_0 A_i^a + \sum_{\mu < \nu}{p^{\mu \nu
}}^a\partial_0 \Phi_{\mu \nu}^a +(\partial_0\eta^a_i)
\overline{\rho}^a_i + (\partial_0\overline{\eta}^a_i) \rho^a_i -h
+A_0^aC^a\right) d^3x,
\end{eqnarray}
where the quantities $p_i=p_i^at^a$, $i=1,2,3$ and $p^{\mu \nu
}={p^{\mu \nu }}^at^a$ are introduced as follows
\begin{eqnarray}
p_i=F_{0i}+\frac{m}{2}\Phi_{0i},~{p^{\mu \nu
}}=-\frac{1}{4}D_0\Phi^{\mu
\nu},~\rho_i=-iD_0\eta_i,~\overline{\rho}_i=iD_0\overline{\eta}_i,
\end{eqnarray}
and the functions $h(x)$ and $C(x)=C^a(x)t^a$ are defined by
\begin{eqnarray}
h=\frac{1}{2}((p_i^a-\frac{m}{2}\Phi_{0i}^a)(p_i^a-\frac{m}{2}\Phi_{0i}^a)+\sum_{i<j}F_{ij}^aF_{ij}^a)-
2\sum_{i<j}{p^{ij}}^a{p^{ij}}^a+2{p^{0i}}^a{p^{0i}}^a+ \nonumber \\
+\frac{1}{8}(-\sum_{i<j}(D_k\Phi_{ij})^a(D_k\Phi_{ij})^a+
\label{ham}
(D_k\Phi_{0i})^a(D_k\Phi_{0i})^a)+\frac{m}{2}\sum_{i<j}\Phi_{ij}^aF_{ij}^a+
\\
+i(\overline{\rho}^a_i\rho_i^a+(D_k\overline{\eta}_i)^a(D_k{\eta}_i)^a),\nonumber
\end{eqnarray}
\begin{equation}\label{constr}
C=D_ip_i+\sum_{\mu<\nu}[p^{\mu \nu},\Phi_{\mu
\nu}]+[\rho_i,\overline{\eta}_i]_+ +[\overline{\rho}_i,\eta_i]_+.
\end{equation}
In the formulas above and thereafter $[\cdot,\cdot]_+$ stands for
the anticommutator.

From formulas (\ref{leg}), (\ref{ham}) and (\ref{constr})  we
deduce that the dynamical system described by Lagrangian
(\ref{actloc2}) is a generalized Hamiltonian system with
Hamiltonian $H=\int hd^3x$, the pairs $(A_i^a,p_i^a)$, $(\Phi_{\mu
\nu}^a, {p^{\mu \nu }}^a)$, $(\eta_i^a, \overline{\rho}_i^a)$,
$(\overline{\eta}_i^a,\rho_i^a)$, for $i=1,2,3$, $\mu <\nu$,
$a=1,\ldots,{\rm dim}~\mathfrak g$ are canonical conjugate
coordinates and momenta on the phase space $\Gamma$ of our system,
$A_0^a$ are Lagrange multipliers, and $C^a$ are constraints
generating the gauge action on the phase space.

The (super)Poisson structure on $\Gamma$ has the standard form,
\begin{eqnarray}
\{ A_i^a({\bf x}),p_j^b({\bf
y})\}=-\delta_{ij}\delta^{ab}\delta({\bf x-y}), \{ \Phi_{\mu
\nu}^a({\bf x}), {p^{\gamma \delta }}^b({\bf y})\}=
-\delta_\mu^\gamma\delta_\nu^\delta\delta^{ab}\delta({\bf x-y}), \label{phase} \\
\{ \eta_i^a({\bf x}), \overline{\rho}_j^b({\bf
y})\}_+=\delta_{ij}\delta^{ab}\delta({\bf x-y}), \{
\overline{\eta}_i^a({\bf x}),\rho_j^b({\bf
y})\}_+=\delta_{ij}\delta^{ab}\delta({\bf x-y}),\nonumber \qquad
\quad
\end{eqnarray}
and all the other (super)Poisson brackets of the canonical
variables vanish.

One can also show that the constraints $C^a$ have the following
Poisson brackets
\begin{equation}\label{cc1}
\{ C^a({\bf x}),C^b({\bf y})\}={\rm g}~C^{ab}_c C^c({\bf
x})\delta({\bf x-y}),
\end{equation}
where $C^{ab}_c$ are the structure constants of the Lie algebra
$\mathfrak g$, $[t^a,t^b]=C^{ab}_ct^c$.

Moreover, for any $a$ the Poisson bracket of the Hamiltonian $H$
and of the constraint $C^a$ vanish,
\begin{equation}\label{cc2}
\{H,C^a \}=0.
\end{equation}

Formulas (\ref{cc1}) and (\ref{cc2}) imply that the constraints
$C^a$ are of the first class. Therefore, according to the general
theory of constrained Hamiltonian systems (see Ref. \cite{FS},
Ch.3, \S 2), the generalized Hamiltonian system with first class
constraints described above is equivalent to the associated usual
Hamiltonian system defined on the reduced phase space $\Gamma^*$.
The description of $\Gamma^*$ presented below is similar to that
in case of the Yang-Mills field, and we refer the reader to Ref.
\cite{FS}, Ch. 3, \S 2 for technical details.

Recall that in order to explicitly describe the reduced space one
needs to impose additional subsidiary (gauge fixing) conditions on
the canonical variables. In the Hamiltonian formulation the most
convenient gauge fixing condition is the Coulomb condition,
\begin{equation}\label{col}
\partial_iA_i=0.
\end{equation}

This condition is admissible in the sense that the determinant of
the matrix of Poisson brackets of the components of the constraint
$C$ and of the components of subsidiary condition (\ref{col}) does
not vanish.

The realization of the reduced space $\Gamma^*$ associated to
subsidiary condition (\ref{col}) is a Poisson submanifold in
$\Gamma$ defined by the following equations
\begin{equation}\label{redsp}
\partial_iA_i=0,~C=D_ip_i+\sum_{\mu<\nu}[p^{\mu \nu},\Phi_{\mu
\nu}]+[\rho_i,\overline{\eta}_i]_+
+[\overline{\rho}_i,\eta_i]_+=0,
\end{equation}
and the Hamiltonian of the associated Hamiltonian system on
$\Gamma^*$ is simply the restriction of the original Hamiltonian
$H$ to $\Gamma^*$.

The first equation in (\ref{redsp}) suggests that it is natural to
use the transversal components of the field with spatial
components $A_i$, $i=1,2,3$, their conjugate momenta and the
variables $(\Phi_{\mu \nu}^a, {p^{\mu \nu }}^a)$, $(\eta_i^a,
\overline{\rho}_i^a)$, $(\overline{\eta}_i^a,\rho_i^a)$ as
canonical coordinates on the reduced space $\Gamma^*$.

Indeed, let $e^i({\bf k})$, $i=1,2$ be two arbitrary orthonormal
vectors such that $e^i({\bf k})\cdot {\bf k=0}$ and $e^1(-{\bf
k})=e^2({\bf k})$. Let $u^1({\bf k})=\frac{e^1({\bf k})+e^2({\bf
k})}{\sqrt{2}}$ and $u^2({\bf k})=i\frac{e^1({\bf k})-e^2({\bf
k})}{\sqrt{2}}$ be the complex linear combinations of $e^1$ and
$e^1$ satisfying the property $u^i({\bf k})={u^i}^*(-{\bf k})$.
Then the real coordinates
\begin{equation}\label{tr1}
{A^i}^a({\bf x})=\frac{1}{(2\pi)^3}\int e^{i{\bf k}\cdot({\bf
x-y})}u^i_j({\bf k})A_j^a({\bf y})d^3kd^3y,~i=1,2
\end{equation}
have the following conjugate momenta
\begin{equation}\label{tr2}
{p^i}^a({\bf x})=\frac{1}{(2\pi)^3}\int e^{-i{\bf k}\cdot({\bf
x-y})}u^i_j({\bf k})p_j^a({\bf y})d^3kd^3y,~i=1,2,
\end{equation}
i.e.
$$
\{ {A^i}^a({\bf x}),{p^j}^b({\bf
y})\}=-\delta^{ij}\delta^{ab}\delta({\bf x-y}).
$$
The longitudinal component $A_{\parallel}$ of the vector field
with spatial components $A_i$, $i=1,2,3$,
\begin{equation}\label{along}
A_{\parallel}^a({\bf x})=\frac{i}{(2\pi)^3}\int e^{i{\bf
k}\cdot({\bf x-y})}\frac{k_j}{|{\bf k}|}A_j^a({\bf y})d^3kd^3y,
\end{equation}
vanishes on $\Gamma^*$ while using the second equation in
(\ref{redsp}), and the fact that subsidiary condition (\ref{col})
is admissible, the momenta $p_{\parallel}^a$ conjugate to the
coordinates $A_{\parallel}^a$,
\begin{equation}\label{plong}
p_{\parallel}^a({\bf x})=-\frac{i}{(2\pi)^3}\int e^{-i{\bf
k}\cdot({\bf x-y})}\frac{k_j}{|{\bf k}|}p_j^a({\bf y})d^3kd^3y,
\end{equation}
can be expressed on the reduced space via the canonical variables
$({A^i}^a,{p^i}^a)$ $(\Phi_{\mu \nu}^a, {p^{\mu \nu }}^a)$,
$(\eta_i^a, \overline{\rho}_i^a)$,
$(\overline{\eta}_i^a,\rho_i^a)$ introduced on $\Gamma^*$ above.
The canonical variables on $\Gamma^*$ are the true dynamical
variables for the system described by Lagrangian (\ref{actloc2}).

\section{The classical unperturbed system}\label{unp}

In Sections \ref{unp} -- \ref{unit} we assume that the mass
parameter $m$ is not equal to zero.

In this section we investigate the classical unperturbed system
for which the coupling constant g vanishes, and the corresponding
equations of motion become linear. For the unperturbed system the
Lagrangian is given by the following formula
\begin{equation}
L_0=\int  \textrm{tr} \left(
\frac{1}{2}F_{\mu\nu}F^{\mu\nu}-\frac{1}{8}(\square\Phi_{\mu\nu})\Phi^{\mu\nu}
+\frac{m}{2}\Phi_{\mu\nu}F^{\mu\nu} \label{actloco}
-2i\sum_{i=1}^3 \overline{\eta}_i (\square \eta_i)\right)  d^3x,
\end{equation}
where
$$
F_{\mu\nu}=\partial_\mu A_\nu-\partial_\nu A_\mu.
$$

In view of the discussion in Section \ref{sect1} we expect that in
a physical sector the unperturbed system describes the massive
vector field with values in the Lie algebra $\mathfrak g$.

For convenience we introduce the following notation for the
components $\Phi_{\mu \nu}^a$ of the ghost field $\Phi_{\mu \nu}$
and for their conjugate momenta ${p^{\mu \nu}}^a$:
\begin{eqnarray*}
G_k^a=\frac{1}{2}\varepsilon_{ijk}\Phi_{ij}^a,~\phi_k^a=\Phi_{0k}^a,\\
P_k^a=\frac{1}{2}\varepsilon_{ijk}{p^{ij}}^a,~\pi_k^a={p^{0k}}^a,
\end{eqnarray*}
and define $\mathfrak g$--valued vector fields $A$, $p$, $G$, $P$,
$\phi$, $\pi$ on ${\mathbb{R}}^3$. By definition these vector
fields have spatial components $A_i$, $p_i$, $G_i$, $P_i$,
$\phi_i$, $\pi_i$, respectively. We also write $E$ for the
$\mathfrak g$--valued vector field on ${\mathbb{R}}^3$, with
components
$$
E_k^a=\frac{1}{2}\varepsilon_{ijk}{F^{ij}}^a.
$$

Let $H_0$ be the free Hamiltonian corresponding to $H$, i.e. $H_0$
is obtained from $H$ by putting g$=0$. Using the new notation one
can rewrite $H_0$ in the following form:
\begin{eqnarray}
H_0=\int d^3x\{\frac{1}{2}((p^a-\frac{m}{2}\phi^a)\cdot
(p^a-\frac{m}{2}\phi^a)+E^a\cdot E^a)-
2P^a\cdot P^a+2\pi^a\cdot \pi^a+  \nonumber \\
 +\frac{1}{8}( G^a \cdot \triangle G^a-
 \phi^a\cdot  \triangle \phi^a)+\frac{m}{2}G^a\cdot E^a+ \label{hamo}
\\
+i(\overline{\rho}^a_i\rho_i^a+(D_k\overline{\eta}_i)^a(D_k{\eta}_i)^a)\}
,\nonumber
\end{eqnarray}
where $\triangle=\partial_i\partial_i$ is the Laplace operator.

For the unperturbed system the constraint $C=0$ is reduced to
\begin{equation}\label{constro}
C=\partial_ip_i=0,
\end{equation}
and the subsidiary condition remains the same,
\begin{equation}\label{consto}
\partial_iA_i=0.
\end{equation}

Clearly, the reduced space $\Gamma_0^*$ associated to constraints
(\ref{constro}) and subsidiary conditions (\ref{consto}) is
isomorphic to $\Gamma^*$.

To describe the dynamics generated by Hamiltonian (\ref{hamo}) on
$\Gamma_0^*$ we shall use the canonical coordinates on $\Gamma^*$
introduced in Section \ref{hamform}. One can further simplify the
study of the equations of motion generated by $H_0$ on the reduced
phase space $\Gamma^*_0$ by introducing the longitudinal and the
transversal components of the coordinates and of the momenta. The
longitudinal components $G_\parallel$ and $\phi_\parallel$ of $G$
and $\phi$ are defined by formulas similar to (\ref{along}) with
$A$ replaced by $G$ and $\phi$, respectively, and the longitudinal
components $P_\parallel$ and $\pi_\parallel$ of $P$ and $\pi$,
which are the conjugate momenta to $G_\parallel$ and
$\phi_\parallel$, are introduced by formulas similar to
(\ref{plong}). By definition the transversal component $A_\perp$
of $A$ is equal to $A-{\rm grad}\triangle^{-1}\partial_iA_i$,
\begin{equation}\label{transv}
A_\perp=A-{\rm grad}\triangle^{-1}\partial_iA_i,
\end{equation}
and the transversal components of other bosonic canonical
variables are defined by formulas similar to (\ref{transv}).

Now the restriction of $H_0$ to the reduced space $\Gamma^*_0$ can
be represented as follows
\begin{eqnarray}
H_0=\int
d^3x\{\frac{1}{2}((p^a_\perp-\frac{m}{2}\phi^a_\perp)\cdot
(p^a_\perp-\frac{m}{2}\phi^a_\perp)+E^a\cdot E^a)- 2P^a_\perp\cdot
P^a_\perp+2\pi^a_\perp\cdot \pi^a_\perp+  \quad \nonumber \\
+\frac{1}{8}( G^a_\perp \cdot \triangle G^a_\perp-
\phi^a_\perp\cdot  \triangle
\phi^a_\perp)+\frac{m}{2}G^a_\perp\cdot E^a- 2P^a_\parallel
P^a_\parallel+\frac{1}{8} G^a_\parallel \triangle G^a_\parallel+
\quad \label{hamor}
\\
+2\pi^a_\parallel \pi^a_\parallel+ \frac{1}{8}\phi^a_\parallel
(-\triangle +m^2)\phi^a_\parallel
+i(\overline{\rho}^a_i\rho_i^a+(D_k\overline{\eta}_i)^a(D_k{\eta}_i)^a)\}.\quad
\nonumber
\end{eqnarray}
Note that in the expression above the transversal and the
longitudinal components are completely separated.

To investigate the dynamics generated by Hamiltonian (\ref{hamor})
we observe that the expression in the r.h.s. of (\ref{hamor}) is
quadratic in canonical variables. Therefore, according to the
general theory of normal forms for quadratic Hamiltonians (see
Ref. \cite{AR}, Appendix 6), Hamiltonian (\ref{hamor}) can be
reduced to a canonical form by a linear symplectic transformation.

Indeed, if we introduce new variables $\overline A_\perp$,
${\overline p}_\perp$, $q_1$, $r_1$, $q_2$, $r_2$,
\begin{eqnarray}
r_1=\sqrt{2}(P_\perp+\frac{1}{4}~{\rm curl}~\phi_\perp), \qquad
\qquad \qquad \qquad \qquad
 \qquad \qquad \qquad \qquad \qquad ~ ~ \nonumber \\
q_1=\sqrt{2}\frac{3m^2-2\triangle}{8m^2}G_\perp-\sqrt{2}\frac{m^2-2\triangle}{4m\triangle}~{\rm
curl}~A_\perp+\sqrt{2}\frac{m^2+2\triangle}{2m^2\triangle}~{\rm
curl}~\pi_\perp,\quad ~  \nonumber \\
r_2=\sqrt{2}(\pi_\perp+\frac{1}{4}~{\rm
curl}~G_\perp-\frac{1}{2}mA_\perp), \qquad \qquad \qquad \qquad \qquad \qquad \qquad \qquad  ~ \label{newcan} \\
q_2=\frac{\sqrt{2}}{2}\frac{2\triangle-m^2}{4m^2}\phi_\perp+\frac{\sqrt{2}}{m}p_\perp+
\frac{\sqrt{2}}{2}\frac{m^2-2\triangle}{m^2\triangle}~{\rm
curl}~P_\perp, \qquad \qquad \qquad \qquad \nonumber \\
\overline p_\perp=p_\perp-\frac{2}{m}~{\rm
curl}~P_\perp+\frac{\triangle-m^2}{2m}\phi_\perp, \qquad \qquad \qquad \qquad \qquad \qquad \qquad \qquad \nonumber \\
\overline A_\perp=\frac{1}{2m}~{\rm
curl}~G_\perp+\frac{2}{m}\pi_\perp , \qquad \qquad \qquad \qquad
\qquad \qquad \qquad \qquad \qquad \qquad  ~ \nonumber
\end{eqnarray}
then the pairs of their components, $({\overline A^i}^a,
{{\overline p}^i}^a)$, $({q_1^i}^a, {r_1^i}^a)$, $({q_2^i}^a,
{r_2^i}^a)$, defined by formulas similar to (\ref{tr1}),
(\ref{tr2}), and the pairs $(\phi_\parallel^a, \pi_\parallel^a)$,
$(G_\parallel^a, P_\parallel^a)$, $(\eta^a_i, \overline
\rho^a_i)$, $(\overline \eta^a_i, \rho^a_i)$ are canonical
conjugate coordinates and momenta on the reduced phase space
$\Gamma_0^*$. Moreover, in terms of the new variables the
Hamiltonian $H_0$ takes the canonical form
\begin{eqnarray}
H_0=\int d^3x\{\frac{1}{2}(\overline p^a_\perp\cdot \overline
p^a_\perp+\overline A_\perp^a\cdot(-\triangle+m^2)\overline
A_\perp^a)-\frac{1}{2}(r_1^a\cdot r_1^a+r_2^a\cdot r_2^a)+
\nonumber \\
+r_1^a\cdot~{\rm curl}~q_2^a-r_2^a\cdot~{\rm
curl}~q_1^a+2\pi^a_\parallel \pi^a_\parallel+
\frac{1}{8}\phi^a_\parallel (-\triangle +m^2)\phi^a_\parallel-
\label{hamor1}
\\
- 2P^a_\parallel P^a_\parallel+\frac{1}{8} G^a_\parallel \triangle
G^a_\parallel
+i(\overline{\rho}^a_i\rho_i^a+(D_k\overline{\eta}_i)^a(D_k{\eta}_i)^a)\}.\nonumber
\end{eqnarray}

Note that Hamiltonian (\ref{hamor1}) and the momenta $r_1$ and
$r_2$ have the following Poisson brackets
\begin{equation}\label{constreq}
\{H_0,r_1\}={\rm curl}~r_2,~\{H_0,r_2\}=-{\rm curl}~r_1,
\end{equation}
and hence $r_1$ and $r_2$ can be regarded as first class
constraints. Therefore recalling the general scheme of constrained
reduction (see Ref. \cite{FS}, Ch.3, \S 2) one can further reduce
the effective number of degrees of freedom using first class
constraints
\begin{equation}\label{constr1}
r_1=0,~r_2=0.
\end{equation}

Since $r_1$, $r_2$ are the momenta conjugate to coordinates $q_1$
and $q_2$ the subsidiary conditions
\begin{equation}\label{subs1}
q_1=0,~q_2=0
\end{equation}
are admissible for constraints (\ref{constr1}), i.e. the
determinant of the matrix of Poisson brackets of the components of
the constraints and of the components of the subsidiary conditions
does not vanish. The reduced space $\Gamma_0^{**}$ associated to
constraints (\ref{constr1}) and subsidiary conditions
(\ref{subs1}) is defined by the following equations in
$\Gamma_0^{*}$
$$
r_1=0,~r_2=0,~q_1=0,~q_2=0,
$$
and the components ${\overline A^i}^a$, ${{\overline p}^i}^a$ of
the transversal parts $\overline A_\perp$, ${\overline p}_\perp$,
the longitudinal components $G_\parallel^a$, $P_\parallel^a$,
$\phi_\parallel^a$ $\pi_\parallel^a$ and $\eta_i^a,
\overline{\rho}_i^a$, $\overline{\eta}_i^a,\rho_i^a$ are canonical
variables on $\Gamma_0^{**}$. We denote by $H_0^r$ the Hamiltonian
$H_0$ restricted to $\Gamma_0^{**}$, $H_0^r=H_0|_{\Gamma_0^{**}}$,
\begin{eqnarray}
H_0^r=\int d^3x\{\frac{1}{2}({\overline p}^a_\perp\cdot {\overline
p}^a_\perp+\overline A^a_\perp\cdot (-\triangle +m^2)\overline
A^a_\perp) +2\pi^a_\parallel \pi^a_\parallel+
\frac{1}{8}\phi^a_\parallel
(-\triangle +m^2)\phi^a_\parallel- \quad \label{hamor2}\\
- 2P^a_\parallel P^a_\parallel+\frac{1}{8} G^a_\parallel \triangle
G^a_\parallel
+i(\overline{\rho}^a_i\rho_i^a+(D_k\overline{\eta}_i)^a(D_k{\eta}_i)^a)\}.\quad
\nonumber
\end{eqnarray}

The equations of motion generated by Hamiltonian (\ref{hamor2})
read
\begin{eqnarray}
\square G_\parallel=0,~\square
\eta_i=0,~\square\overline{\eta}_i=0, \label{eq1} \qquad \qquad \qquad \qquad \qquad \qquad \qquad \qquad \\
(\square +m^2)\phi_\parallel=0, \label{eq3} \qquad \qquad \qquad \qquad \qquad \qquad \qquad \qquad \qquad \qquad  \quad\\
(\square +m^2) \overline A_\perp=0 .\qquad \qquad \qquad \qquad
\qquad \qquad \qquad \qquad \qquad \qquad \label{eq2}
\end{eqnarray}

Therefore the Hamiltonian $H_0^r$ effectively describes
propagation of the two massive transversal components of the field
$\overline A_\perp$, the massive longitudinal component of the
field $\phi$, the massless longitudinal component of $G$ and the
massless fermions $\eta_i^a,~\overline{\eta}_i^a$.

For the purposes of quantization it is natural to introduce the
holomorphic representation for these fields,
\begin{eqnarray}
\overline A_\perp^a({\bf
x})=\frac{1}{(2\pi)^{\frac{3}{2}}}\int\frac{d^3k}{\sqrt{2({\bf
k}^2+m^2)^{\frac{1}{2}}}}\sum_{i=1,2}(b^a_i({\bf k})e^i({\bf
k})e^{i{\bf k}\cdot
{\bf x}}+{b^a_i}^*({\bf k})e^i({\bf k})e^{-i{\bf k}\cdot {\bf x}}), \label{gp2} \qquad \\
\overline p_\perp^a({\bf x})=\frac{i}{(2\pi)^{\frac{3}{2}}}\int
{d^3k}\sqrt{\frac{({\bf
k}^2+m^2)^{\frac{1}{2}}}{2}}\sum_{i=1,2}(-b^a_i({\bf k})e^i({\bf
k})e^{i{\bf k}\cdot {\bf x}}+{b^a_i}^*({\bf k})e^i({\bf
k})e^{-i{\bf k}\cdot {\bf x}}), \nonumber
\end{eqnarray}
\begin{eqnarray}
\eta_i^a({\bf
x})=-\frac{1}{(2\pi)^{\frac{3}{2}}}\int\frac{d^3k}{\sqrt{2|{\bf
k}|}}(c_i^a({\bf k})e^{-i{\bf k}\cdot
{\bf x}}+{c_i^a}^*({\bf k})e^{i{\bf k}\cdot {\bf x}}), \nonumber \quad \\
\overline{\eta}_i^a({\bf
x})=\frac{i}{(2\pi)^{\frac{3}{2}}}\int{d^3k}\frac{d^3k}{\sqrt{2|{\bf
k}|}}(\overline{c}_i^a({\bf k})e^{-i{\bf k}\cdot
{\bf x}}-{\overline{c}^a_i}^*({\bf k})e^{i{\bf k}\cdot {\bf x}}), \label{cran} \\
\rho_i^a({\bf x})=\frac{1}{(2\pi)^{\frac{3}{2}}}\int
{d^3k}\sqrt{\frac{|{\bf k}|}{2}}(c_i^a({\bf k})e^{-i{\bf k}\cdot
{\bf x}}-{c_i^a}^*({\bf k})e^{i{\bf k}\cdot {\bf x}}), \nonumber ~\\
\overline{\rho}_i^a({\bf
x})=\frac{i}{(2\pi)^{\frac{3}{2}}}\int{d^3k}\sqrt{\frac{|{\bf
k}|}{2}}(\overline{c}_i^a({\bf k})e^{-i{\bf k}\cdot {\bf
x}}+{\overline{c}^a_i}^*({\bf k})e^{i{\bf k}\cdot {\bf x}}),
\nonumber ~
\end{eqnarray}
\begin{eqnarray}
G_\parallel^a({\bf
x})=\frac{2}{(2\pi)^{\frac{3}{2}}}\int\frac{d^3k}{\sqrt{2|{\bf
k}|}}(a^a({\bf k})e^{-i{\bf k}\cdot
{\bf x}}+{a^a}^*({\bf k})e^{i{\bf k}\cdot {\bf x}}), \label{gp}\qquad \\
P_\parallel^a({\bf x})=\frac{i}{2(2\pi)^{\frac{3}{2}}}\int
{d^3k}\sqrt{\frac{|{\bf k}|}{2}}(-a^a({\bf k})e^{-i{\bf k}\cdot
{\bf x}}+{a^a}^*({\bf k})e^{i{\bf k}\cdot {\bf x}}), \nonumber
\end{eqnarray}
\begin{eqnarray}
\phi_\parallel^a({\bf
x})=\frac{2}{(2\pi)^{\frac{3}{2}}}\int\frac{d^3k}{\sqrt{2({\bf
k}^2+m^2)^{\frac{1}{2}}}}(b^a({\bf k})e^{i{\bf k}\cdot
{\bf x}}+{b^a}^*({\bf k})e^{-i{\bf k}\cdot {\bf x}}), \label{gp1} \qquad \\
\pi_\parallel^a({\bf x})=\frac{i}{2(2\pi)^{\frac{3}{2}}}\int
{d^3k}\sqrt{\frac{({\bf k}^2+m^2)^{\frac{1}{2}}}{2}}(-b^a({\bf
k})e^{i{\bf k}\cdot {\bf x}}+{b^a}^*({\bf k})e^{-i{\bf k}\cdot
{\bf x}}). \nonumber
\end{eqnarray}
The new complex coordinates $b_i^a$, ${b_i^a}^*$, $c_i^a$,
${\overline{c}_i^a}^*$, $\overline{c}_i^a$, ${c_i^a}^*$, $a^a$,
${a^a}^*$, $b^a$, ${b^a}^*$ have the standard (super)Poisson
brackets,
\begin{eqnarray}
\{ a^a ({\bf k}),{a^b}^* ({\bf k}') \} =i\delta^{ab}\delta({\bf
k-k}'), \qquad \qquad \qquad \qquad \qquad \qquad \qquad \qquad \qquad \nonumber \\
\{b_i^a ({\bf k}),{{b}_j^b}^* ({\bf k}')\}
=i\delta_{ij}\delta^{ab}\delta({\bf k-k}') ,~ \{ b^a({\bf
k}),{b^b}^*({\bf k}') \}
=i\delta^{ab}\delta({\bf k-k}'), \label{pbr} \qquad ~ \\
\{c_i^a ({\bf k}),{\overline{c}_j^b}^* ({\bf k}')\}_+
=i\delta_{ij}\delta^{ab}\delta({\bf k-k}'),~
\{\overline{c}_i^a({\bf k}),{c_j^b}^*({\bf k}')\}_{+}
=i\delta_{ij}\delta^{ab}\delta({\bf k-k}'). \nonumber
\end{eqnarray}

One can express Hamiltonian (\ref{hamor2}) in terms of the
holomorphic coordinates,
\begin{equation}\label{energy}
H_0^r=
 \int d^3k {({\bf k}^2+m^2)^{\frac{1}{2}}}({b^a}^*b^a+{b^a_i}^*b^a_i) -\int
d^3k |{\bf
k}|({\overline{c}^a_i}^*c_i^a-\overline{c}_i^a{c_i^a}^*+{a^a}^*
a^a).
\end{equation}

Now Hamiltonian (\ref{hamor2}) can be naturally split into two
parts,
\begin{equation}\label{split}
H_0^r=H_0^++H_0^-.
\end{equation}

The first one,
\begin{eqnarray}\label{hpp}
H_0^+= \label{h1} \int d^3k {({\bf
k}^2+m^2)^{\frac{1}{2}}}({b^a}^*b^a+{b^a_i}^*b^a_i),
\end{eqnarray}
corresponding to the first  line in formula (\ref{hamor2})
describes propagation of the two transversal massive components of
the gauge field $\overline A_\perp$ and the massive longitudinal
component of the field $\phi$. According to formula (\ref{hpp})
they propagate with positive energy (positive energy sector).
These three components can be regarded as three independent
components of one massive vector field with values in the Lie
algebra $\mathfrak g$ (see Ref. \cite{IZ}, Sect. 3-2-3 for the
description of the dynamics of the massive vector field). We
conclude that in the positive energy sector we have obtained the
desired result: the Hamiltonian $H_0^+$ describes the massive
vector field with values in the Lie algebra $\mathfrak g$.

The second part $H_0^-$,
\begin{equation}\label{hmm}
H_0^-=-\int d^3k |{\bf
k}|({\overline{c}^a_i}^*c_i^a-\overline{c}_i^a{c_i^a}^*+{a^a}^*
a^a),
\end{equation}
corresponding to the last line in (\ref{hamor2}) describes the
massless fields (negative energy sector).  Note that according to
formula (\ref{hmm}) the massless fields indeed propagate with
negative energy (for fermions it is true in the quantum case).

An important property of decomposition (\ref{split}) is that the
positive and the negative energy sectors are Poincar\'{e}
invariant.

Indeed, consider the coordinate transformation which is inverse to
(\ref{newcan}),
\begin{eqnarray}
A_\perp=\overline A_\perp-\frac{\sqrt{2}}{m}r_2, \nonumber \qquad \qquad \qquad \qquad \qquad \qquad ~\\
p_\perp=-\frac{m^2-2\triangle}{2\sqrt{2}\triangle}~{\rm
curl}~r_1+\frac{m\sqrt{2}}{2}q_2, \qquad \qquad \quad \nonumber \\
G_\perp=\sqrt{2}q_1-\frac{2~{\rm curl}}{m}\overline
A_\perp-\frac{-2\triangle+m^2}{\sqrt{2}m^2\triangle}~{\rm
curl}~r_2, ~~\label{inv2} \\
P_\perp=-\frac{\sqrt{2}}{4}~{\rm curl}~q_2+\frac{1}{2m}~{\rm
curl}~\overline p_\perp+\frac{3m^2-2\triangle}{4\sqrt{2}m^2}r_1,
\nonumber \\
\phi_\perp=\sqrt{2}q_2-\frac{2}{m}\overline
p_\perp-\frac{m^2+2\triangle}{m^2\sqrt{2}\triangle}~{\rm
curl}~r_1, \qquad \qquad ~\nonumber \\
\pi_\perp=-\frac{m^2-2\triangle}{4\sqrt{2}m^2}r_2-\frac{\sqrt{2}}{4}~{\rm
curl}~q_1+\frac{m^2-\triangle}{2m}\overline A_\perp.~~\nonumber
\end{eqnarray}

The transformation induced by (\ref{inv2}) on the reduced space
$\Gamma_0^{**}$ has the following form
\begin{eqnarray}
A_\perp=\overline A_\perp, \nonumber \qquad  \quad ~~\\
p_\perp=0, \qquad \qquad ~~ \nonumber \\
G_\perp=-\frac{2~{\rm curl}}{m}\overline
A_\perp, \label{inv1} \\
P_\perp=\frac{1}{2m}~{\rm curl}~\overline p_\perp,
\nonumber \\
\phi_\perp=-\frac{2}{m}\overline
p_\perp,  \qquad \nonumber \\
\pi_\perp=\frac{m^2-\triangle}{2m}\overline A_\perp.\nonumber
\end{eqnarray}

Note that the equations of motion generated by Hamiltonian $H_0^r$
imply that $\overline p_\perp=\frac{\partial \overline
A_\perp}{\partial t}$.

Now observe that the positive energy solutions to equations of
motion (\ref{eq1})--(\ref{eq2}) are the only solutions which
satisfy the Poincar\'{e} invariant condition
\begin{equation}\label{cond1}
d\Phi=0,
\end{equation}
where $\Phi=\Phi_{\mu\nu}dx^\mu\wedge dx^\nu$ is the differential
form with components $\Phi_{\mu\nu}$, and $d$ is the exterior
differential defined on the Minkowski space. Indeed, condition
(\ref{cond1}) takes the following form in components
\begin{equation}\label{c1}
-\partial_0G+{\rm curl}~\phi=0,~\partial_iG_i=0.
\end{equation}
Using formulas (\ref{inv1}) and recalling equations (\ref{eq3}),
(\ref{eq2}) one checks directly that for the positive energy
solutions conditions (\ref{c1}) are satisfied. Therefore the
positive energy sector is Poincar\'{e} invariant.

The condition dual to (\ref{cond1}) with respect to the scalar
product $<\cdot,\cdot>$ of $\mathfrak g$--valued forms on the
Minkowski space,
\begin{equation}\label{scal}
<\Phi,\Psi>=\int d^4x\textrm{tr}(\Phi\wedge *\Psi),
\end{equation}
is
\begin{equation}\label{cond2}
d^*\Phi=0,
\end{equation}
where $*$ is the Hodge star operator associated to the standard
metric on the Minkowski space, and $d^*$ is the operator conjugate
to $d$ with respect to scalar product (\ref{scal}).

Condition (\ref{cond2}) takes the following form in components:
\begin{equation}\label{constr3}
\partial_0\phi=-{\rm curl}~G,~\partial_i\phi_i=0.
\end{equation}
Using the definition of the component $G_\parallel$ and
reconstructing the vector field $\widehat G_\parallel$ on
${\mathbb{R}}^3$, which corresponds to the solution $G_\parallel$,
by the formula
$$
\widehat G_\parallel=-{\rm
grad}~\frac{1}{\sqrt{|-\triangle|}}G_\parallel
$$
one immediately obtains that $\widehat G_\parallel$ is the only
solution to equation of motions (\ref{eq1})--(\ref{eq2}) that
obeys Poincar\'{e} invariant condition (\ref{constr3}). Finally
observe that the fermionic part of Lagrangian (\ref{actloco}) is
obviously Poincar\'{e} invariant. Therefore the negative energy
sector is Poincar\'{e} invariant as well. The unwanted negative
energy sector can be easily split off in the quantum case.

\section{Quantization}\label{SQ}

We start by discussing quantization procedure for the unperturbed
system defined on the phase space $\Gamma^*_0$. To construct the
quantized unperturbed system we shall use the following
coordinates on $\Gamma^*_0$: the holomorphic coordinates $c_i^a$,
${\overline{c}_i^a}^*$, $\overline{c}_i^a$, ${c_i^a}^*$, $a^a$,
${a^a}^*$, $b^a$, ${b^a}^*$, $b^a_i$, ${b^a_i}^*$, and the
components ${q_1^i}^a$, ${r_1^i}^a$, ${q_2^i}^a$, ${r_2^i}^a$ of
$q_1$, $r_1$, $q_2$, $r_2$. After quantization these variables
become operators obeying the standard (super)commutation
relations,
\begin{eqnarray}
~[{\bf a}^a({\bf k}),{{\bf a}^b}^*({\bf
k}')]=\delta^{ab}\delta({\bf k-k}'), \qquad \qquad \qquad \qquad
\qquad \qquad
\qquad \qquad \quad \nonumber \\
~[{\bf b}^a({\bf k}),{{\bf b}^b}^*({\bf
k}')]=\delta^{ab}\delta({\bf k-k}'),~ [{\bf b}_i^a({\bf k}),{{\bf
b}^b_j}^*({\bf k}')]=\delta_{ij}\delta^{ab}\delta({\bf
k-k}'), \qquad  ~  \label{comm} \\
~[{\bf c}_i^a ({\bf k}),{\overline{{\bf c}}_j^b}^* ({\bf k}')]_+
=\delta_{ij}\delta^{ab}\delta({\bf k-k}'),~ [\overline{{\bf
c}}_i^a({\bf k}),{{\bf c}_j^b}^*({\bf k}')]_{+}
=\delta_{ij}\delta^{ab}\delta({\bf
k-k}'), \nonumber \\
~[ {{\bf q}_1^i}^a({\bf x}),{{\bf r}_1^j}^b({\bf
y})]=i\delta^{ij}\delta^{ab}\delta({\bf x-y}),~ [{{\bf
q}_2^i}^a({\bf x}),
{{\bf r}_2^j}^b({\bf y})]= i\delta^{ij}\delta^{ab}\delta({\bf x-y}). \quad \nonumber \\
\end{eqnarray}

We shall use the standard coordinate representation ${\mathcal
H}_Q$ for the operators ${{\bf q}_1^i}^a$, ${{\bf r}_1^i}^a$,
${{\bf q}_2^i}^a$, ${{\bf r}_2^i}^a$. This representation is
diagonal for ${{\bf q}_1^i}^a$ and for ${{\bf q}_2^i}^a$. The
operators ${{\bf c}_i^a}^*$, ${\overline{{\bf c}}^a_i}^*$, ${\bf
c}_i^a$, $\overline{{\bf c}}_i^a$, ${\bf a}^a$, ${{\bf a}^a}^*$,
${\bf b}^a$, ${{\bf b}^a}^*$, ${\bf b}_i^a$, ${{\bf b}_i^a}^*$ act
as usual in the fermionic and in the bosonic Fock spaces,
respectively; all the operators with superscript $*$ being
regarded as creation operators. We denote the Hilbert space tensor
product of the fermionic and of the bosonic Fock spaces by
${\mathcal H}_F$.

We shall also denote by ${{\bf q}_1}$, ${{\bf r}_1}$, ${{\bf
q}_2}$, ${{\bf r}_2}$ the vector valued operator quantities with
components ${{\bf q}_1^i}^a$, ${{\bf r}_1^i}^a$, ${{\bf
q}_2^i}^a$, ${{\bf r}_2^i}^a$.

Let $\mathcal H$ be the Hilbert space tensor product of the
coordinate representation space ${\mathcal H}_Q$ and of the Fock
space ${\mathcal H}_F$, ${\mathcal H}={\mathcal H}_Q
\bigotimes{\mathcal H}_F$. $\mathcal H$ is the space of states for
the quantized system associated to Hamiltonian (\ref{hamor1}).

We overemphasize that the quantized Hamiltonian $H_0$ is a
selfadjoint operator ${\bf H}_0$ acting in the Hilbert space
$\mathcal H$ equipped with a positive definite sesquilinear scalar
product. But the quantization procedure described above does not
guarantee that the energy spectrum of ${\bf H}_0$ belongs to the
positive semiaxis! The negative energy states have, of course, no
physical meaning.

Now recall that actually we need to quantize the system associated
to the reduced Hamiltonian (\ref{hamor2}). According to Dirac's
quantum constraint reduction scheme the quantized Hamiltonian
(\ref{hamor2}) acts in the space ${\mathcal H}_{\rm red}^0$ which
can be obtained from ${\mathcal H}$ by imposing the constraints
${\bf r}_1$ and ${\bf r}_2$,
\begin{equation}\label{redsp0}
{\mathcal H}_{\rm red}^0=\{ |v> \in {\mathcal H}:~{{\bf r}_1^i}^a
|v>=0,~{{\bf r}_2^i}^a |v>=0 \}.
\end{equation}
Note that by construction ${\mathcal H}_{\rm red}^0\simeq
{\mathcal H}_F$.

Now we can define the space of the {\em physical} states
${\mathcal H}_{\rm phys}^0$ for the reduced Hamiltonian
(\ref{hamor2}) by removing the unwanted Poincar\'{e} invariant
negative energy sector. As we observed in the end of the last
section the Poincar\'{e} invariant negative energy sector for the
reduced free Hamiltonian $H_0^r$ contains all the fermions and the
longitudinal component $G_\parallel$ of the spatial part of the
field $\Phi_{\mu\nu}$. Therefore, in view of formulas
(\ref{cran}), (\ref{gp}) and (\ref{comm}), ${\mathcal H}_{\rm
phys}^0$ can be naturally defined as the subspace of ${\mathcal
H}_{\rm red}^0$ which does not contain states with excitations
created by the operators ${{\bf c}_i^a}^*$, ${\overline{{\bf
c}}^a_i}^*$, ${{\bf a}^a}^*$. In other words
\begin{equation}\label{hop}
{\mathcal H}_{\rm phys}^0=\{ |v> \in {\mathcal H}_{\rm
red}^0:~{{\bf c}_i^a}|v>= \overline{{\bf c}}^a_i |v>={{\bf
a}^a}|v>=0 \}.
\end{equation}

Note that the space of the physical states ${\mathcal H}_{\rm
phys}^0$ can also be described with the help of the quantized
Hamiltonian $H_0^-$. Following our convention we denote the
quantized Hamiltonian $H_0^-$ by ${\bf H}_0^-$. From formula
(\ref{hmm}) it immediately follows that
\begin{equation}\label{hop1}
{\mathcal H}_{\rm phys}^0=\{ |v> \in {\mathcal H}_{\rm red}^0:{\bf
H}_0^-|v>=0 \}.
\end{equation}
Since the negative energy sector is Poincar\'{e} invariant the
operator ${\bf H}_0^-$ commutes with the Hamiltonian ${\bf H}_0$,
\begin{equation}\label{hmconstr}
[{\bf H}_0^-,{\bf H}_0]=0.
\end{equation}
At the classical level this can be seen directly from definitions
(\ref{energy}) and (\ref{hmm}). Since $H_0^-$ does not depend on
the canonical variables ${q_1^i}^a$, ${r_1^i}^a$, ${q_2^i}^a$,
${r_2^i}^a$ we also have
$$
[{{\bf r}_1^i}^a,{\bf H}_0^-]=[{{\bf r}_2^i}^a,{\bf H}_0^-]=0.
$$
The last formula together with (\ref{hmconstr}) implies that ${\bf
r}_{1,2}$ and ${\bf H}_0^-$ can be regarded as a system of first
class quantum constraints, and description (\ref{redsp0}),
(\ref{hop1}) of the space of physical states is a realization of
Dirac's quantum constraint reduction scheme.

From the definition of the space ${\mathcal H}$, the description
(\ref{hop}) of the physical subspace and the definition of the
coordinate representation for the operators ${{\bf q}_1^i}^a$,
${{\bf r}_1^i}^a$, ${{\bf q}_2^i}^a$, ${{\bf r}_2^i}^a$ it follows
that ${\mathcal H}_{\rm phys}^0$ is simply the bosonic Fock space
for the operators ${\bf b}^a$, ${{\bf b}^a}^*$, ${\bf b}_i^a$,
${{\bf b}_i^a}^*$, all the operators with superscript $*$ being
regarded as creation operators.

The quantized Hamiltonian ${\bf H}_0$ restricted to ${\mathcal
H}_{\rm phys}^0$ is the quantization of $H_0^+$,
\begin{eqnarray}
{\bf H}_0^+= \int d^3k {({\bf k}^2+m^2)^{\frac{1}{2}}}({{\bf
b}^a}^*{\bf b}^a+{{\bf b}^a_i}^*{\bf b}^a_i). \nonumber
\end{eqnarray}
Therefore the quantized Hamiltonian ${\bf H}_0$ restricted to
${\mathcal H}_{\rm phys}^0$ can be identified with that of the
quantized massive $\mathfrak g$--valued vector field.

\section{Another massive nonabelian theory}\label{unit}

In this section we study the relation of the theory with
unperturbed Lagrangian (\ref{actloco}) and the abelian version of
the theory suggested in Refs. \cite{BCS,HL,L} for describing
massive gauge fields. We show that these two massive theories are
equivalent in the physical sector. Note that in the nonabelian
case the Lagrangian introduced in Refs. \cite{BCS,HL,L} generates
a unitary renormalizable quantum field theory describing the
$\mathfrak g$--valued massive vector field only.

First we recall the definition of the gauge invariant action
introduced in Refs. \cite{BCS,HL,L}. We use the same notation for
gauge fields as in the previous sections. Let $B_{\mu \nu}$ and
$C_\mu$ are the $(2,0)$-type skew-symmetric tensor field and
vector field, respectively, with values in the adjoint
representation of $\mathfrak g$. The action defined in Refs.
\cite{BCS,HL,L} can be written in the following form
\begin{equation}\label{act2}
W=\int d^4x ~\textrm{tr} \Big(
\frac{1}{2}F_{\mu\nu}F^{\mu\nu}-\frac{1}{6}H_{\mu\nu\lambda}H^{\mu\nu\lambda}
-\frac{m}{2}\epsilon^{\mu \nu \rho \lambda}B_{\mu \nu}F_{\rho
\lambda} \Big),
\end{equation}
where $m$ ia a mass parameter,
$$
H_{\mu\lambda\nu}=D_{[\mu}B'_{\lambda \nu]}=D_{[\mu}B_{\lambda
\nu]}+\textrm{g}~[F_{[\mu \lambda},C_{\nu]}],~B'_{\lambda
\nu}=B_{\lambda \nu}-D_{[\lambda}C_{\nu]},
$$
and for the lower indexes square brackets always mean
antisymmetrization, e.g.
$$
D_{[\mu}B'_{\lambda \nu]}=D_{\mu}B'_{\lambda
\nu}+D_{\lambda}B'_{\nu \mu}+D_{\nu}B'_{\lambda \mu},~
D_{[\lambda}C_{\nu]}=D_{\lambda}C_{\nu}-D_{\nu}C_{\lambda}.
$$
$\epsilon^{\mu \nu \rho \lambda}$ is the absolutely antisymmetric
tensor of rank four such that $\epsilon^{0123}=1$.

The action (\ref{act2}) is invariant under the gauge
transformations
\begin{equation}\label{gaugen1}
A_\mu \mapsto \frac{1}{\textrm{g}}(\partial_\mu g)g^{-1}+gA_\mu
g^{-1},~B_{\mu \nu}\mapsto g B_{\mu \nu} g^{-1},~C_\mu \mapsto
gC_\mu g^{-1}
\end{equation}
and under vector transformations
\begin{equation}\label{vtransf}
A_\mu \mapsto A_\mu,~B_{\mu \nu}\mapsto B_{\mu
\nu}+D_{[\lambda}\Lambda_{\nu]},~C_\mu \mapsto C_\mu+\Lambda_\mu,
\end{equation}
where $\Lambda_\mu$ is an arbitrary vector field with values in
the adjoint representation of the gauge group.

In particular, definition (\ref{act2}) and formulas
(\ref{vtransf}) imply that the field $C_\mu$ is not dynamical and
can be removed by transformations (\ref{vtransf}) (see Refs.
\cite{BCS,L4,HL,L} for more detailed discussion of this
phenomenon). In Ref. \cite{L4}, Sect. IV it is also shown that
action (\ref{act2}) describes a massive $\mathfrak g$--valued
vector field.

A precise analysis of the reduced phase space in the framework of
Hamiltonian reduction can be found in Refs. \cite{L4,LL}. Actually
due to the presence of the non--dynamical vector field $C_\mu$ the
explicit description of the reduced phase associated to action
(\ref{act2}) is much more complicated than in the corresponding
abelian case, i.e. when the coupling constant g vanishes, and
hence the axillary vector field $C_\mu$ is not present in the
definition of the action. It turns out that beside of symmetries
(\ref{gaugen1}) and (\ref{vtransf}) action (\ref{act2}) also has
some other hidden symmetries which reduce the number of functional
degrees of freedom of the system to three, like in the
corresponding abelian case when the axillary field $C_\mu$ is not
present (see Refs. \cite{L4,LL}).

In Refs. \cite{HL,L1} it is shown that a BRST invariant
tree--level action can be constructed for the theory introduced in
Refs. \cite{BCS,HL,L}. Therefore the theory is unitary in the
physical sector. Note that the nonabelian massive theory
introduced in Refs. \cite{BCS,HL,L} contains more field variables
than the corresponding abelian theory with ${\rm g}=0$. As a
result one can circumvent the no-go theorem (see Ref. \cite{H})
and prove that the nonabelian theory is renormalizable (see Ref.
\cite{L2}).

Now we consider the abelian counterpart $W_0$ of action
(\ref{act2}) obtained by putting ${\rm g}=0$ in (\ref{act2}),
\begin{equation}\label{act3}
W_0=\int d^4x ~\textrm{tr} \Big(
\frac{1}{2}F_{\mu\nu}F^{\mu\nu}-\frac{1}{6}H_{\mu\nu\lambda}H^{\mu\nu\lambda}
-\frac{m}{2}\epsilon^{\mu \nu \rho \lambda}B_{\mu \nu}F_{\rho
\lambda} \Big),
\end{equation}
where now $F_{\mu\nu}=\partial_\mu A_\nu -
\partial_\nu A_\mu$, and
$$
H_{\mu\lambda\nu}=\partial_{[\mu}B_{\lambda \nu]}.
$$

The action (\ref{act3}) is invariant under the abelian gauge
transformations
\begin{equation}\label{gaugen2}
A_\mu \mapsto A_\mu +\partial_\mu \chi,~B_{\mu \nu}\mapsto B_{\mu
\nu} ,
\end{equation}
and under vector transformations
\begin{equation}\label{vtransf2}
A_\mu \mapsto A_\mu,~B_{\mu \nu}\mapsto B_{\mu
\nu}+\partial_{[\lambda}\Lambda_{\nu]},
\end{equation}
where now $\chi$ and $\Lambda_\mu$ are arbitrary $\mathfrak
g$--valued function and vector field, respectively.

The easiest way to prove that action (\ref{act3}) describes the
massive vector field is as follows. Introducing a vector field
$K^\mu=\frac{1}{2}\epsilon^{\mu \nu \rho
\lambda}\partial_{\nu}B_{\rho \lambda}$ and integrating by parts
one can rewrite $W_0$ in the form
\begin{equation}\label{so}
W_0=\int d^4x ~\textrm{tr} \Big(
\frac{1}{2}F_{\mu\nu}F^{\mu\nu}+K_{\mu}K^{\mu}
-2{m}K^{\mu}A_{\mu}\Big),~F_{\mu\nu}=\partial_\mu A_\nu -
\partial_\nu A_\mu.
\end{equation}
This action gives the following equation of motion for $K_\mu$
$$
K_\mu=mA_\mu.
$$
Substituting $K_\mu$ given by the last formula into (\ref{so}) we
obtain the usual action for the massive $\mathfrak g$-valued
vector field of mass $m$,
\begin{equation}\label{sm}
W_0=\int d^4x ~\textrm{tr} \Big(
\frac{1}{2}F_{\mu\nu}F^{\mu\nu}-m^2A_{\mu}A^{\mu}\Big),~F_{\mu\nu}=\partial_\mu
A_\nu -
\partial_\nu A_\mu.
\end{equation}

In Refs. \cite{HL,L1} it is shown that the physical sector for
action (\ref{act3}) describing the massive vector field can be
defined with the help of the BRST cohomology corresponding to
symmetry transformations (\ref{gaugen2}) and (\ref{vtransf2}). We
show that the corresponding gauge fixed BRST invariant action is
equal to a gauge fixed abelian version of action (\ref{act1}).
This proves that in the abelian case the theory introduced in
Refs. \cite{C,S} is equivalent to that defined in Refs.
\cite{BCS,HL,L}.

First following Refs. \cite{HL,L1} we introduce the ghosts and
gauge fixing conditions corresponding to transformations
(\ref{gaugen2}) and (\ref{vtransf2}). We choose the gauge fixing
terms for these transformations as in Refs. \cite{HL,L1},
\begin{equation}\label{gf}
{\mathcal F}_1=\partial^\mu A_\mu,~{\mathcal F}_2^\mu=\partial_\nu
B^{\mu\nu}.
\end{equation}
The $\mathfrak g$--valued anticommuting ghosts, $\omega,~\overline
\omega$, $\omega^*=\overline \omega$, corresponding to
transformation (\ref{gaugen2}) are introduced in the standard way.
For transformation (\ref{vtransf2}) we note that the r.h.s. of
formula (\ref{vtransf2}) only depends on the transversal part of
the vector field $\Lambda_\mu$. Therefore for transformation
(\ref{vtransf2}) only three pairs of $\mathfrak g$--valued
anticommuting  ghosts, $\omega_i,~\overline \omega_i$,
$\omega_i^*=\overline \omega_i$, $i=1,2,3$, corresponding to the
three components of the transversal part of the vector field
$\Lambda_\mu$ are required. This observation slightly simplifies
the definition of the BRST cohomology comparing to the original
definition given in Refs. \cite{HL,L1}. In order to define the
corresponding BRST transformation we introduce three arbitrary
complex--valued orthonormal vectors $v^1(k),~v^2(k),~v^3(k)$
orthogonal to the position vector $k$ in the Fourier dual to the
Minkowski space. We shall also assume that these vectors satisfy
the following conditions $v^i(k)={v^i}^*(-k)$. Let
\begin{equation}
{\omega}_\mu({ x})=\frac{1}{(2\pi)^4}\int e^{i{ k}\cdot({
x-y})}v^i_\mu({ k})\omega_i({ y})d^4kd^4y,~\mu=0,1,2,3,
\end{equation}
and
\begin{equation}
\overline{\omega}_\mu({ x})=\frac{1}{(2\pi)^4}\int e^{i{ k}\cdot({
x-y})}v^i_\mu({ k})\overline \omega_i({ y})d^4kd^4y,~\mu=0,1,2,3
\end{equation}
be the transversal vector combinations of the ghost fields,
${\omega}^*_\mu=\overline{\omega}_\mu$. The BRST transformation
$\delta$ corresponding to gauge fixing conditions (\ref{gf}) has
the form
\begin{eqnarray}
\delta A_\mu=\partial_\mu \omega\delta\lambda,~\delta
B_{\mu\nu}=\partial_{[\mu}\omega_{\nu]}\delta\lambda, \nonumber \\
\delta \omega=0,~\delta \overline \omega=\partial^\mu A_\mu\delta\lambda, \label{BRST} \\
\delta\omega_\mu=0,~\delta\overline \omega_\mu=\partial^\nu
B_{\mu\nu}\delta\lambda, \nonumber
\end{eqnarray}
where $\delta\lambda$ is an anticommuting parameter independent of
the space--time coordinates. Note that transformation (\ref{BRST})
is in agreement with the transversality condition for the vector
$\overline \omega_\mu$.

To obtain the BRST--invarant action one has to add to action
(\ref{act3}) a term $W_0^{gf}$ containing gauge fixing conditions
(\ref{gf}) and the corresponding Faddeev--Popov operator,
\begin{equation}\label{}
W_0^{gf}=\int d^4x{\rm tr}\big( \partial^\mu A_\mu\partial^\nu
A_\nu-\partial^\nu B_{\mu\nu}\partial_\lambda
B^{\mu\lambda}-2\overline \omega \square
\omega-2\sum_{i=1}^3\overline \omega_i \square \omega_i \big).
\end{equation}

In Refs. \cite{HL,L1} it is proved the gauge fixed action
$W'_0=W_0+W_0^{gf}$,
\begin{eqnarray}\label{actgf}
W'_0=\int d^4x ~\textrm{tr} \Big(
\frac{1}{2}F_{\mu\nu}F^{\mu\nu}-\frac{1}{2}B_{\mu\nu}\square
B^{\mu\nu} -\frac{m}{2}\epsilon^{\mu \nu \rho \lambda}B_{\mu
\nu}F_{\rho \lambda}+\partial^\mu A_\mu\partial^\nu
A_\nu-\\
-2\overline \omega \square \omega-2\sum_{i=1}^3\overline \omega_i
\square \omega_i \Big),\nonumber
\end{eqnarray}
is invariant under BRST transformation (\ref{BRST}). Therefore the
physical sector for the model describing the massive vector field
can be defined by quantizing gauge fixed action (\ref{actgf}) and
by taking the corresponding BRST cohomology.

Introducing the new variables
$\Phi^{\mu\nu}=-\epsilon^{\mu\nu\lambda \rho}B_{\lambda \rho}$ and
$\eta=\frac{\omega+\overline \omega}{\sqrt{2}}$, $\overline
\eta=\frac{\omega-\overline \omega}{\sqrt{2i}}$,
$\eta_i=\frac{\omega_i+\overline \omega_i}{\sqrt{2}}$, $\overline
\eta_i=\frac{\omega_i-\overline \omega_i}{\sqrt{2i}}$ one can
rewrite action (\ref{actgf}) in the form
\begin{eqnarray}\label{actloc20}
W'_0=\int  \textrm{tr} \left(
\frac{1}{2}F_{\mu\nu}F^{\mu\nu}-\frac{1}{8}(\square
\Phi_{\mu\nu})\Phi^{\mu\nu}
+\frac{m}{2}\Phi_{\mu\nu}F^{\mu\nu}-2i\sum_{i=1}^3
\overline{\eta}_i (\square \eta_i)+ \right.\\
\left. +\partial^\mu A_\mu
\partial^\nu A_\nu-2i\overline{\eta} (
\square \eta)\right) d^4x.\nonumber
\end{eqnarray}
Action (\ref{actloc20}) coincides with the abelian counterpart of
action (\ref{act1}) with the additional gauge fixing and
Faddeev--Popov terms for the Lorentz gauge $\partial^\mu A_\mu=0$.
Thus the theory with Lagrangian (\ref{actloco}) is equivalent to
the theory with action (\ref{act3}) in the physical sector. In
that sector both theories describe the massive $\mathfrak
g$--valued vector field.

\section{Conclusion}

As we observed in this paper one can suggest at least two mass
generation mechanisms for gauge fields. They correspond to
different quadratic terms for the $(2.0)$--type tensor field. In
the abelian case these mass generation mechanisms are equivalent.
The problem of equivalence of the two theories in the nonabelian
case is still open. This question can be studied in the framework
of BRST cohomology. If the two theories are equivalent then there
should exist a gauge fixing term and a set of ghosts corresponding
to symmetry (\ref{vtransf}) for action (\ref{act2}) such that the
corresponding BRST--invariant gauge fixed action defined as in
Refs. \cite{HL,L1} coincides with (\ref{act1}).

Actually one can construct other nonabelian gauge invariant
actions which differ from (\ref{act1}) or (\ref{act2}) by terms
quadratic in the $(2.0)$--type tensor field. An interesting
related question is which actions defined in this way generate
unitary renormalizable theories describing a massive vector field
only? At present it is only known that the theory suggested in
Refs. \cite{BCS,HL,L} has all these properties.

Another interesting problem is: can the gauge invariant actions
with mass terms mentioned above be generated dynamically in a
nonperturbative way when we consider initially massless models? If
such a possibility was realized one could obtain a dynamical gauge
invariant generation mechanism of a mass parameter in a gauge
theory.

\vskip 0.5cm

\section*{Acknowledgments}

The author is grateful to Amitabha Lahiri for a useful discussion.

\end{document}